\documentclass{optica-article}
\journal{opticajournal} 

\articletype{Research Article}
\usepackage{graphicx}
\usepackage{dsfont}
\usepackage{physics}

\newcommand{\revise}[1]{{#1}}
\newcommand{\reviseII}[1]{{#1}}

\begin{document}

\title{Noise-robust discrimination of incoherent point sources with spatial-mode demultiplexing}

\author{Jian-Qiang Liu,\authormark{1} Chao-Ning Hu,\authormark{1} Jun Xin, \authormark{1,*} and Xiao-Ming Lu\authormark{1,$\dagger$}}

\address{\authormark{1}School of Physics and Zhejiang Key Laboratory of Quantum State Control and Optical Field Manipulation, Hangzhou Dianzi University, Hangzhou 310018, China}

\email{\authormark{*}jxin@hdu.edu.cn}
\email{\authormark{$\dagger$}lxm@hdu.edu.cn}

\begin{abstract*}
    We theoretically predict and experimentally demonstrate that a reduced spatial-mode demultiplexing (SPADE) measurement using only the two lowest-order Hermite-Gaussian modes exhibits remarkable robustness against background noise in discriminating between a single source and two incoherent point sources.
    We establish a theoretical framework incorporating uniform background noise and derive an analytical Chernoff exponent expression, showing that SPADE consistently outperforms direct imaging (DI) across the sub-Rayleigh separations .
    Experimental results confirm that SPADE-based hypothesis testing approaches the quantum limit even when the background-to-signal photon ratio per pixel is 0.11.
    This advantage stems from SPADE's ability to concentrate source information into minimal detection modes, reducing the cumulative background noise impact.
    Our findings provide a practical detection scheme for applications where background noise is inevitable, such as astronomical observations and quantum sensing.
\end{abstract*}

\section{Introduction}

A central problem in optical imaging is the discrimination of closely spaced incoherent point sources, e.g., whether an observed signal arises from a single source or from two closely spaced, incoherent point sources~\cite{Helstrom1973}.
Hypothesis testing provides a rigorous statistical framework for making decisions under uncertainty, which is essential for extracting reliable conclusions from noisy or limited data~\cite{Wasserman2010,Hayashi2005}.
This task is particularly relevant in fields such as astronomy, where resolving individual celestial bodies is often limited by the diffraction of light.
Due to diffraction, the images of two point sources can overlap significantly, and as their separation decreases, the resulting intensity distribution becomes nearly indistinguishable from that of a single source.
This leads to a dramatic decrease in the resolving power of direct imaging, a limitation commonly referred to as ``Rayleigh's curse''~\cite{Rayleigh1879}.

Inspired by quantum detection and estimation theory~\cite{Helstrom1976,Holevo2011}, seminal works have revealed that the root cause of Rayleigh's curse lies in traditional direct intensity measurements, which discard crucial phase information of the optical field in the image plane~\cite{Tsang2016,Nair2016,Lupo2016}.
Remarkably, this limitation can be overcome by employing optimized measurement strategies.
A pioneering approach in this regard is spatial-mode demultiplexing (SPADE)~\cite{Tsang2016}, which measures the projection coefficients of the optical field onto a set of Hermite-Gaussian (HG) modes.
It has been shown that when the separation between two point-like sources falls within the sub-Rayleigh regime, most of the Fisher information about their separation can be captured by a few detectors coupled to one or several spatial modes~\cite{Tsang2016,Nair2016b,Tham2017,Yang2016,Paur2016,Tang2016,Paur2018,Boucher2020,Tan2023,Rouviere2024,Aiello2025,Wallis2025a,paneru2026a,santamaria2023b,deshler2025,gozzard2025}.
Theoretical analyses have shown that the Fisher information about the separation under the truncated SPADE scheme that measures the HG modes up to some order approaches its quantum limit as the order increases~\cite{Rehacek2017a} and can readily surpass the Fisher information of direct imaging (DI) for sub-Rayleigh separations.

Beyond enhancing parameter estimation, SPADE has also been shown to optimize the error probability in hypothesis testing tasks~\cite{Lu2018,PhysRevLett.127.130502,PhysRevLett.129.180502,Schlichtholz2024a,PhysRevA.110.052602,Zhang2024e,Wadood2024,Wallis2025,PhysRevA.111.023706,linowski2025a,jha2025a,shringarpure2026a,dealmeida2021b,Amato2026,Wallis2026}, specifically for discriminating between a single source and two incoherent sources.
Lu {\it et al.}~\cite{Lu2018} showed that the quantum Chernoff bound for this binary discrimination problem can be approached by a binary SPADE scheme distinguishing a specific HG mode from all orthogonal modes.
In the context of asymmetric hypothesis testing, SPADE has been demonstrated to outperform direct imaging by significantly suppressing the false-positive error rate, particularly in applications such as exoplanet detection~\cite{PhysRevLett.127.130502}.
Beyond these fundamental limits, the advantage of SPADE extends to more practical and complex scenarios.
It has been theoretically established that SPADE remains quantum-optimal for discriminating among libraries of arbitrary real-world objects, significantly outperforming direct imaging in identifying complex shapes beyond simple point sources~\cite{PhysRevLett.129.180502}.

We focus on the impact of experimental noise on SPADE's performance in discriminating between a single source and two incoherent sources.
Practical implementations are inevitably subject to various imperfections, including mode crosstalk, background light, and detector dark counts.
For the crosstalk, recent studies have demonstrated that SPADE can maintain its superiority over direct imaging in the presence of realistic mode crosstalk through optimized statistical tests~\cite{Schlichtholz2024a,Wadood2024,Amato2026}.
However, the effect of background noise from the photon counters has not yet been studied.

In this work, we theoretically predict and experimentally demonstrate that a reduced SPADE measurement, utilizing merely the two lowest-order HG modes, can exhibit remarkable robustness against background noise in hypothesis testing tasks.
Assuming uniform background noise across all photon counters, we show that the ability of the SPADE achieves a higher Chernoff exponent compared to direct imaging.
Here, the Chernoff exponent quantifies the exponential decay rate of the error probability with increasing observation time or photon number.
This advantage stems from the ability of the SPADE to efficiently concentrate source information into a minimal number of detector pixels.
In contrast, DI disperses the signal over a large array of pixels; since each pixel contributes independent background noise, the cumulative noise in DI scales with the number of pixels, thereby significantly degrading the discrimination performance.

\section{Theoretical Analysis}

We consider binary hypothesis testing between $H_{\mathrm{I}}$ (a single thermal source) and $H_{\mathrm{II}}$ (two mutually incoherent thermal sources).
Assume that the sources are weak and the optical field is in a quasi-monochromatic regime, allowing us to model the optical field as a collection of independent temporal modes.
For each temporal mode,  the quantum state of the optical field in the image plane can be modeled as a mixture of the vacuum and single-photon components~\cite{Tsang2016}:
\begin{equation}
    \eta_\alpha \approx (1-\epsilon_\alpha) \op{\mathrm{vac}} + \epsilon_\alpha \rho_\alpha + O(\epsilon^2),
\end{equation}
where $\alpha \in \{\mathrm{I},\mathrm{II}\}$ labels the hypothesis $H_\alpha$, $\epsilon_\alpha \ll 1$ is the probability that a photon arrives at the image plane within a temporal mode, $\ket{\mathrm{vac}}$ denotes the vacuum state, and $\rho_\alpha$ is the single-photon state.

The spatial-mode measurement on each temporal mode is described by a positive-operator-valued measure (POVM) $\{E_q\}$ on the single-photon Hilbert space, where each element $E_q$ corresponds to the spatial modes coupled to the \(q\)-th detector.
Such a measurement detects no photon with probability approximately $1-\epsilon_\alpha$, and a photon in the $q$-th detector with probability approximately $\epsilon_\alpha \tr(\rho_\alpha E_q)$.
Assuming that \(M\) independent temporal modes are available per sampling interval, the total state of the optical field can be expressed as $\eta_\alpha^{\otimes M}$.
Based on the measurement outcomes, one can perform a decision to choose either $H_{\mathrm{I}}$ or $H_{\mathrm{II}}$.
The performance of the decision is characterized by the error probability, which is defined as the probability of making an incorrect decision.
The minimum error probability over all possible decision strategies for the given measurement strategy indicated by the superscript \(\mathrm{(meas)}\) asymptotically decreases with $M$ as $P_\mathrm{e,min}^\mathrm{(meas)} \sim \exp(- M \xi^{\mathrm{(meas)}})$, where \(\xi^{\mathrm{(meas)}}\) is the Chernoff exponent~\cite{chernoff1952a,VanTrees2013} defined as
\begin{align}
    \xi^{\mathrm{(meas)}}
    = -\log \min_{0 \le s \le 1}
    \qty[
        (1-\epsilon_\mathrm{I})^s (1-\epsilon_\mathrm{II})^{1-s}
        + \epsilon_\mathrm{I}^s \epsilon_\mathrm{II}^{1-s} \sum_q p_{\mathrm{I},q}^s p_{\mathrm{II},q}^{1-s}
    ]
\end{align}
with \(p_{\alpha,q} = \tr(\rho_\alpha E_q)\) being the probability of detecting a photon in the \(q\)-th detector under hypothesis \(H_\alpha\).

For realistic imaging systems, the number of temporal modes $M$ is typically large and the mean photon number per temporal mode $\epsilon_\alpha$ is very small.
It is convenient to take the Poisson limit by letting $M \to \infty$ and $\epsilon_\alpha \to 0$ while keeping the average photon number per sampling $\nu_\alpha = M \epsilon_\alpha$ fixed.
In this Poisson limit, the classical Chernoff exponent obeys~\cite{tsang2021a}
\begin{align}
    \lim_{M\to\infty} M \xi^{\mathrm{(meas)}}
    = \max_{0 \le s \le 1} \left[
        s \nu_{\mathrm{I}} + (1-s) \nu_{\mathrm{II}}
        - \sum_q \nu_{\mathrm{I},q}^{\,s} \nu_{\mathrm{II},q}^{\,1-s}
    \right], \label{eq:classical_chernoff}
\end{align}
where $\nu_{\alpha,q} = \nu_\alpha p_{\alpha,q} $ is the mean photon number at the $q$-th detector.

To study the robustness against noise, we consider excess noise, from background light and detector dark counts, contaminating the measurement outcomes of the detectors.
Assuming that the excess noise is Poissonian and uniform across all detectors, the mean photon number at the \(q\)-th detector per sampling is
\begin{align}
    u_{\alpha,q} = \nu_{\alpha,q} + b,
\end{align}
where \(\nu_{\alpha,q}\) is the mean number of the signal photons under the hypothesis \(H_\alpha\) and \(b\) is the mean number of noise photons.
In this work, we consider the scenario where the mean numbers of the signal photons under the two hypotheses are equal, i.e., \(\nu_\mathrm{I} = \nu_\mathrm{II}=\nu\).
Taking \(\nu\) as the resource for the discrimination task, we define the classical error exponent in the Poisson limit as
\begin{align} \label{eq:chernoff_exponent_def}
    \xi_\mathrm{P}^{\mathrm{(meas)}} \equiv \frac1{\nu} \lim_{M\to\infty} M \xi^{\mathrm{(meas)}}.
\end{align}
This means that the minimum error probability under the Poisson limit decays exponentially as \(P_\mathrm{e,min}^\mathrm{(meas)} \sim \exp\,\!(- \nu\,\xi_\mathrm{P}^{\mathrm{(meas)}})\).
Since the sum of two Poisson variables is also Poisson distributed with mean equal to the sum of the individual means, by replacing \(\nu_{\alpha,q}\) with \(u_{\alpha,q}=\nu_{\alpha,q} + b\) in Eq.~(\ref{eq:classical_chernoff}), we obtain
\begin{align} \label{eq:classical_chernoff_equal}
    \xi_\mathrm{P}^{\mathrm{(meas)}}
    = 1 - \min_{0 \le s \le 1} \sum_q
    \qty[
        \qty(p_{\mathrm{I},q} + \frac{b}{\nu})^{\,s} \qty(p_{\mathrm{II},q} + \frac{b}{\nu})^{\,1-s} - \frac{b}{\nu}
    ].
\end{align}
This expression provides a convenient framework for analyzing the performance of different measurement strategies for discriminating between the two hypotheses with equal total mean photon numbers in the presence of uniform excess noise.
As shown in Appendix~\ref{appendix:noise_monotonicity}, the classical Chernoff exponent \(\xi_\mathrm{P}^{\mathrm{(meas)}}\) is a monotonically decreasing function of the excess noise \(b\), indicating that the discrimination performance deteriorates as the excess noise increases.
In the limit of large excess noise, i.e., \(b \gg \nu\), the classical Chernoff exponent approaches zero, indicating that the discrimination task becomes increasingly challenging.

The Chernoff exponent \(\xi_\mathrm{P}^{\mathrm{(meas)}}\) depends on the specific measurement strategy and can be used as a figure of merit to optimize the measurement design.
The maximum of \(\xi_\mathrm{P}^{\mathrm{(meas)}}\) over all possible measurement strategies is given by the quantum Chernoff exponent~\cite{ogawa2004a,kargin2005a,audenaert2007,nussbaum2009a,audenaert2008a}.
For \(N\) detectors each coupled to an orthogonal spatial mode, we obtain the quantum limit of Eq.~\eqref{eq:classical_chernoff_equal} as
\begin{align} \label{eq:quantum_chernoff_equal}
    \xi_\mathrm{P} = 1 -  \min_{0 \le s \le 1}
    \tr[
        \qty(\rho_\mathrm{I} + \frac{b}{\nu} \mathds{1})^s
        \qty(\rho_\mathrm{II} + \frac{b}{\nu} \mathds{1})^{1-s}
        - \frac{b}{\nu} \mathds{1}
    ],
\end{align}
where \(\mathds{1}\) is the identity on the \(N\)-mode Hilbert space.
We provide a detailed derivation of Eq.~\eqref{eq:quantum_chernoff_equal}, as well as a more general discussion, in Appendix~\ref{appendix:quantum_chernoff}.

We now describe the specific source configurations for the two hypotheses.
The density operators of the one-photon states corresponding to the two hypotheses are given by
\begin{align} \label{eq:rho}
    \rho_\mathrm{I}  = \op{\psi}, \qand
    \rho_\mathrm{II} = \frac{\nu_+}{\nu}\op{\psi_+} + \frac{\nu_-}{\nu} \op{\psi_-},
\end{align}
where \(\ket{\psi}\) is the single-photon state under $H_{\mathrm{I}}$, \(\ket{\psi_\pm}\) are the single-photon states under $H_{\mathrm{II}}$ with mean photon numbers \(\nu_\pm\), and \(\nu = \nu_+ + \nu_-\).
Assuming a spatially invariant and diffraction-limited one-dimensional imaging system, the single-photon states are \(\ket{\psi} \equiv \int \dd{x} \psi(x) \ket{x}\) and \(\ket{\psi_\pm} \equiv \int \dd{x} \psi(x \pm d/2) \ket{x}\), where \(\psi(x)\) is the normalized amplitude PSF, \(\ket{x}\) is the position eigenket, and \(d\) is the source separation under $H_{\mathrm{II}}$.
We further assume that the PSF is well approximated by a Gaussian function, given by
\begin{equation} \label{eq:gaussian_PSF}
    \psi(x) = \left(\frac{1}{2\pi\sigma^2}\right)^{1/4} \exp\left(-\frac{x^2}{4\sigma^2}\right)
\end{equation}
with \(\sigma\) being the standard deviation of the PSF.

We here emphasize that the origin of the coordinate system for SPADE is chosen to coincide with the geometric centroid of the two sources under $H_{\mathrm{II}}$.
When the brightnesses of the two sources are unequal, the geometric centroid of the two sources differs from the intensity-weighted centroid that is easier to determine experimentally.
For such cases, additional efforts are required to determine the geometric center, such as the two-point-source PSF fitting method~\cite{Stetson1987,Piatti2016}.

We consider the SPADE measurement and compare it with the quantum limit and DI.
For SPADE, the POVM elements are \(E_q = \op*{\phi_q}\) with \(\ket*{\phi_q}\) denoting the HG modes.
The mode functions of the HG modes are given by
\begin{equation}
    \phi_q(x) = \left(\frac{1}{2\pi\sigma^2}\right)^{1/4} \frac{1}{\sqrt{2^q q!}} H_q\!\left(\frac{x}{\sqrt{2}\sigma}\right) \exp\left(-\frac{x^2}{4\sigma^2}\right),
\end{equation}
where $H_q(\cdot)$ is the $q$-th order Hermite polynomial.
For SPADE with HG modes, the probabilities of a signal photon being detected at the \(q\)-th detector are given by
\begin{align}
    p^\mathrm{(SPADE)}_{\mathrm{I},q}  = \delta_{q0} \qand
    p^\mathrm{(SPADE)}_{\mathrm{II},q} = \frac{1}{q!} \qty(\frac{d}{4\sigma})^{2q} \exp(-\frac{d^2}{16\sigma^2})
\end{align}
for the hypotheses $H_{\mathrm{I}}$ and $H_{\mathrm{II}}$, respectively.
Substituting these probabilities into Eq.~\eqref{eq:classical_chernoff_equal}, we can numerically evaluate the classical Chernoff exponent for SPADE with excess noise.
It is worth noting that for \(d\leq 2\sigma\) (sub-Rayleigh regime), the majority of the signal photons are concentrated in the two lowest-order HG modes as their projection probabilities are larger than 0.97.
This implies that a reduced SPADE measurement, which only measures the two lowest-order HG modes, can capture nearly all the information about the source configuration in the sub-Rayleigh regime.

We shall compare the performance of SPADE with that of DI, which is a standard imaging technique that measures the intensity distribution of the optical field in the image plane.
For DI, the optical field at each pixel with a finite size \(a\) is fed into a photon counter, which suffers from excess noise.
The probabilities of a signal photon being detected at the \(j\)-th pixel are given by
\begin{align}
    p^\mathrm{(DI)}_{\mathrm{I},j}  &= \int_{ja-a/2}^{ja+a/2} |\psi(x)|^2 \dd{x}, \\
    p^\mathrm{(DI)}_{\mathrm{II},j} &= \frac{\nu_+}{\nu} \int_{ja-a/2}^{ja+a/2} \qty|\psi\qty(x+\frac{d}{2})|^2 \dd{x}
    + \frac{\nu_-}{\nu} \int_{ja-a/2}^{ja+a/2} \qty|\psi\qty(x-\frac{d}{2})|^2 \dd{x}
\end{align}
for the hypotheses $H_{\mathrm{I}}$ and $H_{\mathrm{II}}$, respectively.

\begin{figure}[tb]
    \centering
    \includegraphics[]{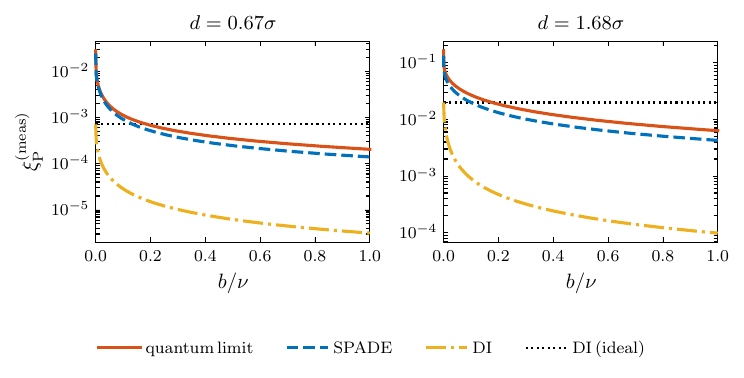}
    \caption{ \label{fig:theoretical_chernoff_exponent}
        Numerical results for the quantum Chernoff exponent (red solid line) and the classical Chernoff exponents for SPADE (blue dashed line) and DI (yellow dash-dotted line) with excess noise.
        Here, the total mean photon number under the two hypotheses is set to be equal, i.e., \(\nu = \nu_+ + \nu_-\).
        The PSF is Gaussian with the standard deviation $\sigma = 115\,\mu\mathrm{m}$.
        The DI is set to 1000 pixels with a pixel size $a = 4.6\,\mu\mathrm{m}$.
        For SPADE, only the two lowest-order HG modes are included in the calculation of the Chernoff exponent.
        The black dotted line shows the classical Chernoff exponent for idealized DI without noise, included as a reference.
    }
\end{figure}

Figure~\ref{fig:theoretical_chernoff_exponent} presents the numerical results for the quantum Chernoff exponent (red solid line) and the classical Chernoff exponents for reduced SPADE and DI with uniform detector noise, for a small separation \(d=0.67\sigma\) and an intermediate separation \(d=1.68\sigma\) in the sub-Rayleigh regime.
The details of the numerical calculations of the quantum Chernoff exponent are provided in Appendix~\ref{appendix:quantum_chernoff}.
Our results reveal that, in terms of the asymptotic error exponent, the SPADE measurement with only the two lowest-order HG modes exhibits superior performance compared to DI when the separation between the two hypothetical sources is in the sub-Rayleigh regime, even in the presence of excess noise.

\begin{figure}[tb]
    \centering
    \includegraphics{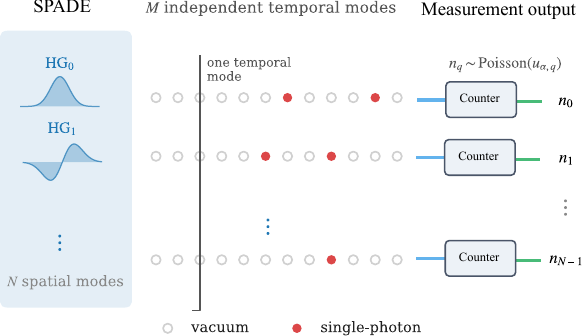}
    \caption{Illustration of the temporal and spatial modes of the optical field in the image plane. The optical field is assumed to be in a weak-source regime, where each temporal mode can be approximated as a mixture of vacuum and single-photon states and the photon number counted by each detector follows a Poisson distribution.}
    \label{fig:spatial-temporal}
\end{figure}

Based on the measurement outcomes, we need a decision rule to determine which hypothesis is true.
A time-resolved record is not necessary, as the photon count \(n_q\) recorded by the \(q\)-th detector is a sufficient statistic~\cite{Wasserman2010} for the mean photon number \(u_{\alpha, q}\) in the Poisson limit, which captures all the information about the distinction between the two hypotheses.
The SPADE scheme is illustrated in Fig.~\ref{fig:spatial-temporal}, where the optical field in the image plane is decomposed into a set of orthogonal spatial modes (only two lowest-order HG modes for the reduced SPADE) and the photon counts \(\mathbf{n} = (n_{0}, n_{1}, \dots, n_{N-1})\) are recorded.
The joint probability mass function of \(\mathbf{n}\) in the weak-source regime and the Poisson limit can be approximated as Poisson statistics~\cite{tsang2021a,smith2010a}:
\begin{equation}\label{eq:Prob}
    P(\mathbf{n} | H_\alpha) = \prod_{q=0}^{N-1} \frac{1}{n_{q}!} (u_{\alpha,q})^{n_{q}} \exp(-u_{\alpha,q}).
\end{equation}
The optimal decision rule is the likelihood ratio test~\cite{VanTrees2013}.
The log-likelihood ratio for this problem is given by
\begin{align} \label{eq:log_likelihood_ratio}
    \log \Lambda(\mathbf{n})
    = \log \frac{P(\mathbf{n} | H_{\mathrm{II}})}{P(\mathbf{n} | H_{\mathrm{I}})}
    = \nu_{\mathrm{I}} - \nu_{\mathrm{II}} +
    \sum_{q=0}^{N-1} \qty(
        n_q \log\frac{p_{\mathrm{II},q} + b/\nu}{p_{\mathrm{I},q} + b/\nu}
    ),
\end{align}
based on which we choose \(H_\mathrm{I}\) if \(\log\Lambda(\mathbf{n})\) is negative and \(H_\mathrm{II}\) if \(\log\Lambda(\mathbf{n})\) is positive.
If \(\log\Lambda(\mathbf{n})\) is zero, we randomly choose either \(H_\mathrm{I}\) or \(H_\mathrm{II}\) with equal probability.
Note that the first two terms in Eq.~\eqref{eq:log_likelihood_ratio} will be canceled out when \(\nu_\mathrm{I} = \nu_\mathrm{II}\), which is the case in our experiment.

\section{Experimental Results}

\begin{figure}[tb]
    \centering
    \includegraphics[width=3.4in]{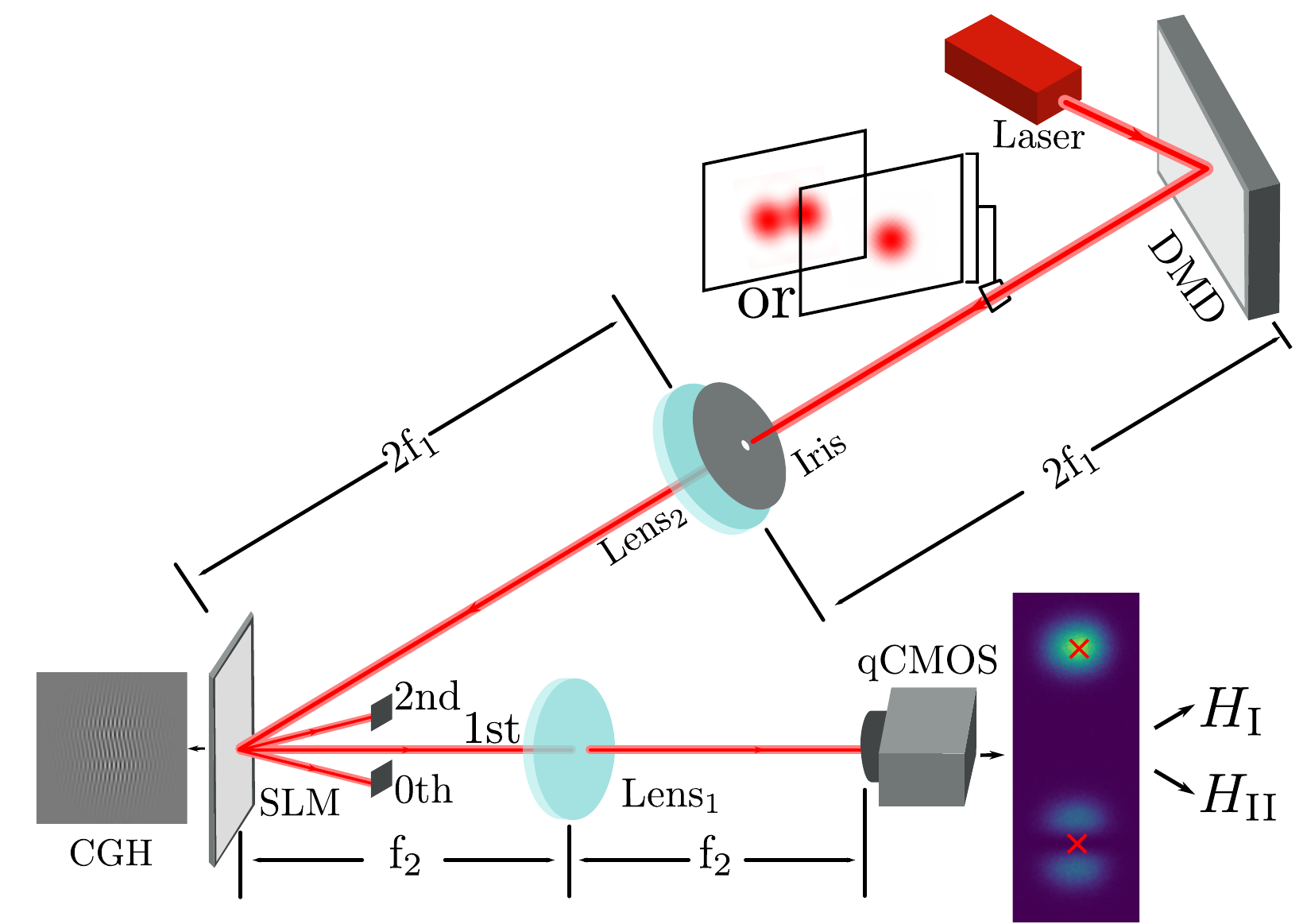}
    \caption{
        Experimental setup for the SPADE-based hypothesis testing.
        DMD, digital micromirror device; CGH, computer-generated hologram; SLM, spatial light modulator.
    Only the first-order diffraction from the SLM is directed to the CMOS camera.}
    \label{fig:setup}
\end{figure}

We experimentally demonstrate the performance superiority of SPADE in hypothesis testing by implementing a reduced SPADE that measures only the two lowest-order HG modes.
The experimental setup is illustrated in Fig.~\ref{fig:setup}.

The source configurations, i.e., a single point source under $H_{\mathrm{I}}$ and two incoherent point sources with approximately equal brightness under $H_{\mathrm{II}}$, are realized using a digital micromirror device (DMD; VIALUX, V-7001 VIS), which is illuminated by a laser beam (Photodigm, 770DBRL-T08) with a wavelength of 770~nm.
For the $H_{\mathrm{I}}$ case, only a single micromirror is switched to the ``on'' state.
To realize the $H_{\mathrm{II}}$ configuration, two selected micromirrors on the DMD are alternately switched at a high frequency of $2\times10^{4}$~Hz.
This switching period is much shorter than the duration of a single sampling window.
As a result, within each sampling interval, the detector effectively integrates light from both positions as an incoherent mixture, thereby emulating two independent incoherent point sources.
After reflection from the DMD, the emitted light passes through an iris and a focusing lens (Lens$_{2}$), forming an Airy disk pattern in the image plane of Lens$_{2}$ with a characteristic width of $\sigma=115~\mu\mathrm{m}$.

We implement the SPADE measurement using a digital holographic technique~\cite{arrizon2007b,rosales-guzman2017a,Paur2016,Hu2025}.
A phase-only spatial light modulator (SLM; HoloEye, PLUTO-2-NIR-011), which displays a computer-generated hologram designed to decompose the two lowest-order HG modes $\ket{\phi_0}$ and $\ket{\phi_1}$, is placed at the image plane.
A CMOS camera (Hamamatsu, ORCA-Quest qCMOS, C15550-20UP) is positioned at the Fourier plane of a second lens (Lens$_{1}$) to measure the corresponding projection intensities.
For each sampling with an exposure time of $50$~ms, the photon numbers $n_0$ and $n_1$ of two specific pixels, marked by red crosses in the inset of Fig.~\ref{fig:setup}, are recorded, corresponding to the two HG modes.

It is worth noting that the holographic modal decomposition method used in our experiment  introduces loss of optical power and is therefore not a perfect implementation of SPADE.
However, this loss is not fundamental; it arises from the use of off-axis holography and spatial-frequency carriers to implement spatial mode decomposition in this experiment~\cite{rosales-guzman2017a}.
The ratio of the optical power in the above-mentioned two specific pixels is equal to the ratio of the projection intensities onto the corresponding HG modes.
Therefore, despite the loss of optical power, the measured ratios still accurately reflect the modal decomposition.
The purpose of this experiment is to provide a proof-of-principle demonstration of the superiority of SPADE in discriminating one versus two point sources, rather than to demonstrate the superiority of the holographic implementation.
Therefore, the loss introduced by the holographic implementation does not affect the validity of our demonstration of SPADE's superiority in hypothesis testing.

\begin{figure}[tb]
    \centering
    \includegraphics[]{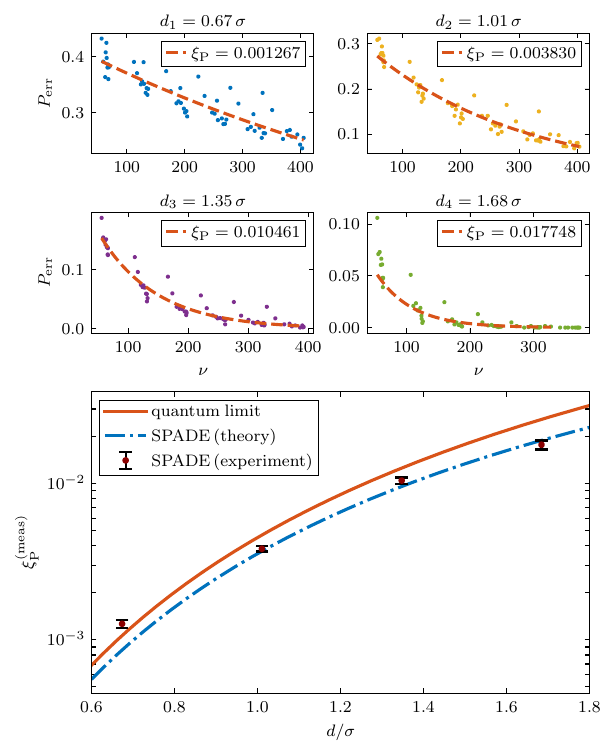}
    \caption{
        Error probability and Chernoff exponent of SPADE-based hypothesis testing.
        The background-to-signal ratio in this experiment is \( b / \nu \approx 0.11\).
        The first four panels show the error probability $P_{\mathrm{err}}$ of SPADE as a function of \(\nu\) for the separations $d_1 = 0.67\sigma$, $d_2 = 1.01\sigma$, $d_3 = 1.35\sigma$, and $d_4 = 1.68\sigma$, respectively.
        The solid lines are fits to $P_{\mathrm{err}} \propto \exp(-\nu \xi_\mathrm{P})$, with $\xi_\mathrm{P}$ the extracted Chernoff exponent.
        The bottom panel shows the extracted Chernoff exponent $\xi_\mathrm{P}$ as a function of the separation $d$.
    }
    \label{fig:experimental_result}
\end{figure}

The excess noise in the experiment is introduced by placing an LED source in front of the CMOS camera, which generates  approximately uniform background light across the entire sensor area~\cite{Hu2025}.
The strength of the background light can be controlled by adjusting the brightness of the LED.
The value of the mean background photon number per detector \(b\) was experimentally determined by measuring the photon counts recorded by the camera in the absence of the signal light source.

We conducted experiments for different separations \(d\) between the two incoherent sources under $H_{\mathrm{II}}$, specifically \(d_1 = 0.67\sigma\), \(d_2 = 1.01\sigma\), \(d_3 = 1.35\sigma\), and \(d_4 = 1.68\sigma\).
At each separation, we collected a total of \(6\times 10^3\) frames for each hypothesis to estimate the error probability, with each frame corresponding to a single physical sampling interval.
From each frame, we extracted the photon counts \(n_0\) and \(n_1\) for the two HG modes.
By averaging the photon counts over all frames, we obtained the mean total photon number \(u=\nu+2b\) per sampling for each separation.
Using the measured mean total photon number \(u\) and the known mean background photon number \(b\), we calculated the value of \(b/\nu = b/(u-2b)\), which is required for the application of the decision rule in Eq.~(\ref{eq:log_likelihood_ratio}).

To estimate the Chernoff exponent, we increased the data frame by frame for the hypothesis testing, which is equivalent to increasing the mean signal photon number per sampling.
Concretely, we used the sum of the photon counts of \(k\) frames as a single sample to perform hypothesis testing for \(k = 1, 2, \dots, 6\).
For each \(k\), we recorded the number of incorrect decisions and calculated the error probability under each hypothesis, based on 1000 independent samples for each hypothesis.

Figure~\ref{fig:experimental_result} presents the experimental results of the error probability $P_{\mathrm{err}}$ as a function of the mean signal photon number $\nu$ for different separations $d$ and the extracted Chernoff exponent $\xi_\mathrm{P}$.
The error probability is obtained by repeating the hypothesis testing for $10^3$ independent samples for both hypotheses and calculating the fraction of incorrect decisions.
We perform a least-squares linear fit on the logarithm of the error probability $P_{\mathrm{err}} \propto \exp(-\nu \xi_\mathrm{P})$ to extract the Chernoff exponent $\xi_\mathrm{P}$, which characterizes the exponential decay rate of the error probability with increasing mean signal photon number $\nu$.
The extracted Chernoff exponent $\xi_\mathrm{P}$ is plotted as a function of the separation $d$ in the bottom panel of Fig.~\ref{fig:experimental_result}.
It agrees well with the theoretical prediction of the classical Chernoff exponent for SPADE with excess noise, as shown by the blue dashed line in Fig.~\ref{fig:theoretical_chernoff_exponent}, and approaches the quantum limit of the Chernoff exponent, as shown by the red solid line in Fig.~\ref{fig:theoretical_chernoff_exponent}.
The experimental results demonstrate that the SPADE-based hypothesis testing is near-optimal and robust against excess noise, even when only the two lowest-order HG modes are measured.

\revise{
    As shown in the bottom panel of Fig.~\ref{fig:experimental_result}, the Chernoff exponent for \(d_1=0.67\sigma\) is above the quantum limit.
    This may be attributed to the pre-asymptotic effects in our experiment, where the finite range of the mean signal photon number $\nu$ does not fully capture the asymptotic behavior required for the Chernoff exponent.
    The Chernoff exponent is defined through the asymptotic decay of the minimum error probability, $P_\mathrm{err} \sim \exp(-\nu \xi_{\text{P}})$, and therefore strictly characterizes the limit $\nu \to \infty$, whereas our experiment can only access a finite range of $\nu$.
    At the separation \(d = 0.67\sigma\) in our experiment, the two hypotheses are the hardest to distinguish: the error probability is the largest and its range of variation is small, and the accessible range of \(\nu\) is not sufficient to fully enter the asymptotic exponential regime.
}

\begin{figure}[htbp]
    \centering
    \includegraphics{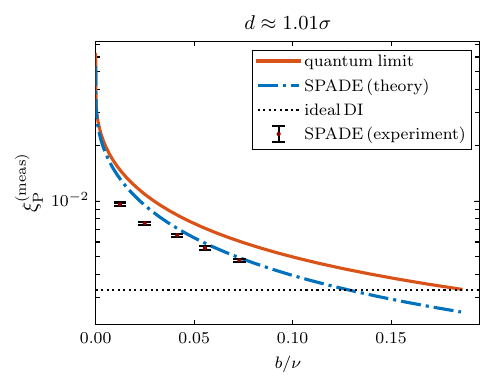}
    \caption{
        Chernoff exponent versus different background noise levels.
        The black dotted line represents the Chernoff exponent for idealized noiseless DI as a reference.
    }
    \label{fig:different_b}
\end{figure}

We also investigated the impact of different background noise levels on the Chernoff exponent, as shown in Fig.~\ref{fig:different_b}.
As the background noise level increases, the Chernoff exponent decreases, indicating a degradation in the hypothesis testing performance.

\reviseII{
    The noise levels examined above can be related quantitatively to concrete applications, since
    \(b/\nu\) is fixed by the ratio of the background surface brightness to the total source
    flux and by the solid angle \(\Omega\) subtended by a detection channel; 
    it is independent of the collecting area, the exposure time, and the overall throughput:
    \begin{equation} \label{eq:bnu_applications}
        \frac{b}{\nu} = \Omega\,10^{-0.4(S-m)} ,
    \end{equation}
    where \(S\) and \(m\) are the surface brightness of the diffuse background and the total magnitude
    of the source in the same band, and \(\Omega\) is the solid angle of the channel.
    This relation follows directly from the flux ratio: the mean background photon number collected
    in a channel of solid angle \(\Omega\) is proportional to the background flux \(\Omega f_b\),
    while \(\nu\) is proportional to the total source flux \(f_s\), so \(b/\nu \propto \Omega f_b/f_s\).
    Expressing the flux ratio in magnitudes, \(f_b/f_s = 10^{-0.4(S-m)}\), yields the equation above.
    For the dark, moonless sky of faint-source astronomy, \(S\simeq21.5\)--\(22~\mathrm{mag\,arcsec^{-2}}\) in the visible~\cite[Table 4]{Leinert1998}, and the near-infrared sky is likewise dominated by zodiacal emission of comparable brightness~\cite[Table 6]{Leinert1998}.
    Taking \(S=21.5~\mathrm{mag\,arcsec^{-2}}\), the value \(b/\nu=0.11\) of our main experiment
    then corresponds to \(m\simeq22.6\) for a \(0.2''\times0.2''\) pixel, to \(m\simeq19.9\) for a
    seeing-limited channel of diameter \(0.8''\), and to \(m\simeq24.9\) for a channel matched to
    the diffraction-limited spot of a 2.4-m telescope at \(800~\mathrm{nm}\), whose solid angle is
    \(\Omega\simeq(\lambda/D)^2=4.7\times10^{-3}~\mathrm{arcsec^2}\).
    The experimental noise levels \(0\le b/\nu\le0.15\) considered here therefore bracket
    the transition to the background-limited regime relevant to the detection of faint
    companions and exoplanets~\cite{PhysRevLett.127.130502}.
}

We remark that although the separation \(d\) is assumed to be known when applying the decision rule in our experiments, the SPADE-based hypothesis testing can be extended to the case of unknown separation by employing a generalized likelihood ratio test~\cite{VanTrees2013}.
It is worth noting that the arrangement of the SPADE measurement is independent of the unknown separation, so that there is no need to use extra photons to first estimate the separation before performing the hypothesis test.
All the separation-dependent information is extracted from the measurement outcomes themselves, without requiring additional photons solely for the purpose of estimating the separation.
After obtaining the separation estimate from the measurement outcomes, it can be directly used in the generalized likelihood ratio test for hypothesis testing.
In other words, the separation estimation and hypothesis testing both utilize the same set of measurement outcomes, ensuring that no extra photons are consumed solely for the purpose of estimating the separation.
This advantage was first recognized in our previous work~\cite{Lu2018}.
Moreover, it is known that the SPADE measurement with only the two lowest-order HG modes retains high sensitivity to the separation even in the sub-Rayleigh regime, allowing for accurate separation estimation.
The separation independence of the SPADE measurement arrangement and the high efficiency of the separation estimation together ensure that the SPADE-based hypothesis testing remains efficient even when the separation under the alternative hypothesis is unknown.

\section{Conclusion}
In this work, we have theoretically predicted and experimentally demonstrated that a reduced SPADE measurement, utilizing only the two lowest-order HG modes, exhibits remarkable robustness against background noise in the binary hypothesis testing task of discriminating between a single source and two incoherent point sources.
Our experiments confirm the theoretically predicted SPADE exponent, which exceeds that of DI in the sub-Rayleigh regime and approaches the quantum limit.
This advantage stems from SPADE's ability to concentrate source information into a minimal number of detection modes, thereby significantly reducing the cumulative impact of background noise.
In contrast, DI disperses the signal across a large array of pixels, where each pixel contributes independent background noise, leading to performance degradation as the number of pixels increases.
Our findings highlight the practical advantage of SPADE in real-world applications where background noise is inevitable.
This advantage becomes particularly pronounced in two scenarios.
First, when the signal is too weak compared to the dark count of the photon detectors, as in low-light astronomical observations where the number of signal photons is inherently limited.
Second, when the background noise is substantial compared to the signal, such as in target discrimination under approximately uniform daytime ambient illumination.
These results thus establish SPADE as a noise-resilient and resource-efficient measurement strategy for practical quantum-limited imaging and sensing tasks.

\appendix

\section{Monotonicity of the classical Chernoff exponent with respect to noise} \label{appendix:noise_monotonicity}

We here prove that the classical Chernoff exponent \(\xi_\mathrm{P}^{\mathrm{(meas)}}\) given in Eq.~(\ref{eq:classical_chernoff_equal}) is a monotonically decreasing function of the noise parameter \(b\).
To do so, we first define the function
\begin{align}
    f(x) = (\alpha + x)^s (\beta+x)^{1-s} - x,
\end{align}
where \(\alpha\) and \(\beta\) are positive constants, \(0 < s < 1\), and \(x \ge 0\).
The derivative of \(f(x)\) with respect to \(x\) is given by
\begin{align}
    f'(x) &= s (\alpha + x)^{s-1} (\beta+x)^{1-s} + (1-s) (\alpha + x)^s (\beta+x)^{-s} - 1 \notag \\
    &=  \frac{s (\beta+x) + (1-s) (\alpha + x) - (\alpha + x)^{1-s} (\beta+x)^{s}}{(\alpha + x)^{1-s} (\beta+x)^{s}}.
\end{align}
Due to the weighted arithmetic-geometric mean inequality~\cite{Hardy1934}, the numerator of the above expression is non-negative.
Therefore, we have \(f'(x) \ge 0\) for all \(x \ge 0\), which implies that \(f(x)\) is a monotonically increasing function of \(x\).

Since each term in the summation of Eq.~(\ref{eq:classical_chernoff_equal}) is of the form \(f(b/\nu)\) with \(\alpha = p_{\mathrm{I},q}\) and \(\beta = p_{\mathrm{II},q}\), and the summation over \(q\) and the minimization over \(s\) preserve the monotonicity, we conclude that \(\xi_\mathrm{P}^{\mathrm{(meas)}}\) is a monotonically decreasing function of \(b\).

\section{Derivation and computation of the quantum Chernoff exponent} \label{appendix:quantum_chernoff}

We first derive the quantum Chernoff exponent \(\xi_\mathrm{P}\) for the binary hypothesis testing problem of discriminating between a single source and two incoherent point sources in the presence of excess noise.
In the Poisson limit, the quantum limit of the Chernoff exponent Eq.~\eqref{eq:classical_chernoff} is given by~\cite{tsang2021a}
\begin{align}
    \lim_{M\to\infty} M \xi^\mathrm{(meas)}
    \leq \max_{0 \le s \le 1} \left[
        s \nu_{\mathrm{I}} + (1-s) \nu_{\mathrm{II}}
        - \tr(\Gamma_{\mathrm{I}}^s \Gamma_{\mathrm{II}}^{1-s})
    \right],
\end{align}
where \(\Gamma_\alpha \equiv \nu_\alpha \rho_\alpha\) are the intensity operators introduced in Ref.~\cite{tsang2021a}.
In the presence of excess noise, which is assumed to be Poissonian and uniform across all detectors, the mean photon number at the \(q\)-th detector per sampling is \(u_{\alpha,q} = \nu_{\alpha,q} + b\).
For \(N\) detectors each coupled to an orthogonal spatial mode, by defining
\begin{align} \label{eq:Gamma_tilde}
    \Gamma_\alpha \to \tilde{\Gamma}_\alpha  = \Gamma_\alpha + b \mathds{1},
\end{align}
we have \(u_{\alpha,q} = \tr(\tilde{\Gamma}_\alpha E_q)\).
By replacing \(\nu_{\alpha,q}\) with \(u_{\alpha,q}=\nu_{\alpha,q} + b\) and \(\Gamma_\alpha\) with \(\tilde\Gamma_\alpha\) in Eq.~(\ref{eq:quantum_chernoff}), we obtain the quantum limit of the Chernoff exponent in the presence of excess noise:
\begin{align} \label{eq:quantum_chernoff}
    \lim_{M\to\infty} M \xi^\mathrm{(meas)}
    \leq \zeta \equiv
    \max_{0 \le s \le 1} \left[
        s u_{\mathrm{I}} + (1-s) u_{\mathrm{II}}
        - \tr(\tilde{\Gamma}_{\mathrm{I}}^s \tilde{\Gamma}_{\mathrm{II}}^{1-s})
    \right],
\end{align}
where \(u_\alpha = \sum_q u_{\alpha, q}\) is the mean total photon number per sampling under hypothesis \(H_\alpha\).
For the case of equal mean number of signal photons, i.e., \(\nu_\mathrm{I} = \nu_\mathrm{II} = \nu\), the above expression leads to the quantum Chernoff exponent used in the main text, i.e., Eq.~(\ref{eq:quantum_chernoff_equal}).

We give a systematic procedure to numerically calculate the quantum Chernoff exponent with Eq.~(\ref{eq:quantum_chernoff}).
To do so, we need to diagonalize the intensity operators with \(\rho_\mathrm{I}\) and \(\rho_\mathrm{II}\) given in Eq.~(\ref{eq:rho}).
Note that \(\tilde\Gamma_\mathrm{I}\) and \(\tilde\Gamma_\mathrm{II}\) differ only on the 3-dimensional subspace spanned by \(\{\ket{\psi}, \ket{\psi_+}, \ket{\psi_-}\}\).
Therefore, \(\tilde\Gamma_\alpha\) can be expressed in the following matrix form:
\begin{align}
    \tilde\Gamma_\alpha = (T_\alpha + b \mathds{1}_3) \oplus b \mathds{1}_{N-3},
\end{align}
where \(\mathds{1}_n\) denotes the \(n\)-dimensional identity matrix and \(T_\alpha\) are the matrix representations of \(\nu\rho_\alpha\) in the 3-dimensional subspace.
Suppose that \(T_\alpha\) has the spectral decomposition
\begin{equation}
    T_\alpha
    = \sum_{j=1}^3 t_{\alpha,j}  \op*{\phi_{\alpha,j}},
\end{equation}
where $\ket*{\phi_{\alpha,j}}$ denotes the eigenbasis of \(T_\alpha\) with the eigenvalues $t_{\alpha,j}$.
We then have
\begin{align}
    \tr(\tilde\Gamma_{\mathrm{I}}^s \tilde\Gamma_{\mathrm{II}}^{1-s})
    = \sum_{j,k=1}^3 (t_{\mathrm{I},j}+b)^s (t_{\mathrm{II},k}+b)^{1-s}
    \abs{\ip{\phi_{\mathrm{I},j}}{\phi_{\mathrm{II},k}}}^2
    + (N-3) b.
\end{align}
Substituting the above expression into Eq.~(\ref{eq:quantum_chernoff}), we obtain
\begin{align} \label{eq:quantum_chernoff_final}
    \zeta
    = \max_{0 \le s \le 1} \big[ 
        s \nu_\mathrm{I} + (1-s) \nu_\mathrm{II} + 3b -   \sum_{j,k=1}^3
        (t_{\mathrm{I},j}+b)^s (t_{\mathrm{II},k}+b)^{1-s}
        \abs{\ip{\phi_{\mathrm{I},j}}{\phi_{\mathrm{II},k}}}^2
    \big].
\end{align}
This formula enables us to numerically evaluate the quantum limit of Chernoff exponent with the eigenvalues and eigenvectors of \(T_\mathrm{I}\) and \(T_\mathrm{II}\).
For the case of equal signal strengths under two hypotheses, we set \(\nu_\mathrm{I} = \nu_\mathrm{II} = \nu\) in Eq.~(\ref{eq:quantum_chernoff_final}), which is equivalent to evaluating Eq.~(\ref{eq:quantum_chernoff_equal}).

For the convenience of numerical calculation, \(T_\mathrm{I}\) and \(T_\mathrm{II}\) should be simultaneously expressed in an orthonormal basis of the 3-dimensional relevant subspace.
This can be done via the Gram matrix \(G_{jk} = \ip*{g_j}{g_k}\) with \(\ket{g_1}=\ket{\psi_0}, \ket{g_2}=\ket{\psi_+}, \ket{g_3}=\ket{\psi_-}\).
Diagonalize the Gram matrix as \(G = U \Lambda U^\dagger\), where \(\Lambda\) is a diagonal matrix with the eigenvalues \(\lambda_j\) of \(G\) and \(U\) is a unitary matrix whose columns are eigenvectors of \(G\).
It is easy to verify that an orthonormal basis \(\{\ket*{e_a}\}\) can be obtained from the non-orthogonal basis \(\{\ket*{g_j}\}\) via the following transformation:
\begin{align}
    \ket*{e_a} = \frac{1}{\sqrt{\lambda_a}} \sum_k  U_{ka} \ket*{g_k}.
\end{align}
For an operator \(A\) expressed with the non-orthogonal basis as \(A = \sum_{j,k} A'_{jk} \op*{g_j}{g_k}\), the matrix representation of \(A\) with respect to the aforementioned orthonormal basis can be calculated as
\begin{align}
    \mel{e_a}{A}{e_b} = \frac{1}{\sqrt{\lambda_a \lambda_b}} \sum_{j,k} U^*_{ja} (G A' G)_{jk} U_{kb},
\end{align}
which in matrix form is equivalent to
\begin{align} \label{eq:matrix_representation}
    A = \Lambda^{1/2} U^\dagger A' U \Lambda^{1/2}.
\end{align}
For the Gaussian PSF given in Eq.~(\ref{eq:gaussian_PSF}), the Gram matrix \(G\) can be simply expressed as
\begin{align} \label{eq:Gram_matrix}
    G = \mqty(
        1 & \delta(\frac{d}{2}) & \delta(\frac{d}{2}) \\
        \delta(\frac{d}{2}) & 1 & \delta(d) \\
        \delta(\frac{d}{2}) & \delta(d) & 1
    ),
\end{align}
where \(\delta(d) \equiv \exp[-d^2/(8\sigma^2)]\) is the overlap between two Gaussian PSFs separated by a distance \(d\).
In terms of Eq.~(\ref{eq:matrix_representation}), the representations of \(\nu \op{\psi}\) and \(\nu_+\op{\psi_+} + \nu_-\op{\psi_-}\) with respect to the non-orthonormal basis \(\{\ket*{g_j}\}\) are given by
\begin{align}
    T'_\mathrm{I} &= \mqty(
        \nu & 0 & 0 \\
        0 & 0 & 0 \\
        0 & 0 & 0
    ) \qand
    T'_\mathrm{II} = \mqty(
        0 & 0 & 0 \\
        0 & \nu_+ & 0 \\
        0 & 0 & \nu_-
    ),
\end{align}
respectively.
Combining the above results and the spectral decomposition of \(G\) given in Eq.~(\ref{eq:Gram_matrix}), we can obtain the 3-dimensional matrix of \(T_\mathrm{I}\) and \(T_\mathrm{II}\), which will be convenient for numerical evaluation of eigenvalues and eigenvectors.\\

\noindent\textbf{Funding.}
The Quantum Science and Technology-National Science and Technology Major Project (Grant No. 2024ZD0301000), the National Natural Science Foundation of China (Grants No. 92476118 and No. 12275062), and Key R\&D Program of Zhejiang (Grant No. 2026C01004).
\\

\noindent\textbf{Disclosures.} The authors declare no conflicts of interest. \\

\noindent\textbf{Data availability.} Data underlying the results presented in this paper are not publicly available at this time but may be obtained from the authors upon reasonable request.

\bibliography{reference}

\end{document}